\documentclass[12pt,preprint]{emulateapj}
\usepackage{apjfonts}
\begin{document}

\title{Transient and Highly Polarized Double-Peaked H$\alpha$ Emission in the\\Seyfert 2 Nucleus of NGC~2110}

\author{Edward C.\ Moran,\altaffilmark{1} Aaron J.\ Barth,\altaffilmark{2}
        Michael Eracleous,\altaffilmark{3} and Laura E.\ Kay\altaffilmark{4}}

\altaffiltext{1}{Astronomy Department, Wesleyan University, Middletown,
                 CT 06459; emoran@wesleyan.edu.}
\altaffiltext{2}{Department of Physics and Astronomy, University of California
                 at Irvine, 4129 Frederick Reines Hall, Irvine, CA 92697-4575;
                 barth@uci.edu}
\altaffiltext{3}{Department of Astronomy \& Astrophysics and Center for
                 Gravitational Wave Physics, The Pennsylvania State University,
                 University Park, PA 16802;
                 mce@astro.psu.edu.}
\altaffiltext{4}{Department of Physics and Astronomy, Barnard College,
                 3009 Broadway, New York, NY 10027; lkay@barnard.edu.}

\begin{abstract}
We have discovered an extremely broad, double-peaked H$\alpha$ emission
line in the polarized flux spectrum of NGC~2110, establishing that this
well-studied Seyfert~2 galaxy contains a disk-like hidden broad-line
region (BLR).  Several properties of NGC 2110 suggest that it is an
obscured twin of Arp 102B, the prototypical double-peaked emission-line
active galactic nucleus (AGN).  A comparison between our data and previous
spectra of NGC~2110 indicates that the
double-peaked H$\alpha$ feature is transient.  The presence of a disk-like
BLR in NGC~2110 has important implications for AGNs: it expands the range
of properties exhibited by Seyfert~2 galaxies, and the fact that the BLR
is obscured by a torus-like structure provides the first evidence that
double-peaked emitters and classical Seyfert nuclei may have the same basic
parsec-scale geometry.

\end{abstract}


\section{Introduction}

NGC~2110, an S0/E galaxy at a distance of $\sim$~30 Mpc, is host to one
of the most extensively studied active galactic nuclei (AGNs) in the local
universe.  The nuclear activity in this object was originally recognized
because of its intense X-ray emission (Bradt et al.\ 1978), and it has
been a popular target for X-ray investigations ever since (Mushotzky 1982;
Turner \& Pounds 1989; Weaver et al.\ 1995; Hayashi et al.\ 1996; Malaguti
et al.\ 1999).  NGC~2110 is classified as a type~2
Seyfert galaxy based on its nuclear optical spectrum, which is dominated
by narrow emission lines (McClintock et al.\ 1979).  Observations of the
the circumnuclear environment of NGC 2110 have revealed peculiar kinematics
and complex structure at all wavelengths (Ulvestad \& Wilson 1983; Wilson
et al.\ 1985; Pogge 1989; Haniff et al.\ 1991; Mulchaey et al.\ 1994;
Ferruit et al.\ 1999, 2004; Quillen et al.\ 1999; Gonz\'alez Delgado et
al.\ 2002; Evans et al.\ 2006)  As a result, NGC~2110 has emerged as an
important laboratory for studying the interaction between an active nucleus
and its surroundings.

NGC~2110 has been frequently considered in the context of AGN unification
(e.g., Antonucci 1993).  The presence of a hidden type~1 nucleus is implied 
by X-ray observations, which indicate heavy absorption of the soft X-ray
spectrum and a Seyfert~1-like intrinsic luminosity at hard X-ray energies
(Mushotzky 1982; Turner \& Pounds 1989; Malaguti et al.\ 1999).  Similarly,
if the extended narrow-line region is powered by photoionization from the
nucleus, the morphology of this region and the strength of the Balmer
recombination lines relative to the observed ionizing flux would suggest
that the nuclear continuum source is obscured from our view (Wilson et al.\
1985; Mulchaey et al.\ 1994).  Yet despite a number of direct searches for
broad emission lines in both the optical (Shuder 1980; Veilleux 1991) and
near-infrared (Veilleux et al.\ 1997; Storchi-Bergmann et al.\ 1999; Knop
et al.\ 2001; Lutz et al.\ 2004), conclusive evidence that NGC~2110 harbors
a hidden broad-line region (BLR) has been elusive.  In this
{\it Letter}, we report the discovery of a spectacular broad H$\alpha$ emission
line in the polarized flux spectrum of this object.

\section{Observations}

We observed NGC~2110 on 30 December 2005 (UT) with the LRIS spectropolarimeter
(Cohen et al.\ 1997) on the Keck~I 10~m telescope.  Both beams of the
instrument were employed in conjunction with the D560 dichroic beam splitter.
On the blue side, a 400 l~mm$^{-1}$ grism blazed at 3400~\AA\ was used,
providing coverage of the 3200--5700~\AA\ range at $\sim 12$~\AA\ resolution
(FWHM).  The red side data, obtained with a 600 l~mm$^{-1}$ grating blazed at
7500~\AA , span 5520--7990~\AA\ at $\sim$~4.5~\AA\ resolution. A
900~s exposure was acquired at each of the four waveplate positions using a
1\farcs0 slit, which was oriented at the mid-observation parallactic angle
of 16$^{\circ}$.  The observations were obtained under photometric conditions
at an average airmass of 1.13.  We extracted one-dimensional spectra from
the reduced images within a 3\farcs15 region centered on the galaxy's nucleus.
Spectropolarimetric processing and measurements were performed following the
methods described by Miller et al.\ (1988) and Barth et al.\ (1999).

Because of the proximity of NGC~2110 to the Galactic plane
($b = -16.\negthinspace \negthinspace^{\circ}5$, $E(B-V)$ = 0.375 mag),
we explored whether
transmission through Galactic dust might affect the observed polarization.
A search of the Heiles (2000) catalog of stellar polarizations reveals
two stars within $\sim 12'$ of NGC~2110 with linear polarization
data. The stars have nearly identical polarization levels and position angles;
using the average values of these ($p = 0.33$\%, $\theta = 34.\negthinspace
\negthinspace^{\circ}5$), we made synthetic $q$ and $u$ spectra assuming a
Serkowski et al.\ (1975) curve for $p(\lambda)$.  These
were then subtracted from the normalized Stokes parameters for NGC~2110 to
correct the galaxy's polarization and position-angle spectra.  As discussed
below, these corrections prove to be relatively minor.

\section{Broad, Polarized H$\alpha$ Emission}

Our spectropolarimetry results are displayed in Figure~1.  
The total-flux spectrum of NGC~2110 ({\it top panel}) is clearly dominated
by narrow emission lines.  The small but significant amount of continuum
polarization observed ({\it middle panel}) is approximately constant with
wavelength, suggesting that it is caused by electron scattering.  In the
5200--6200 \AA\ range the polarization and polarization angle are $p = 0.8$\%
and $\theta = 70^{\circ}$, respectively.  After correction for interstellar
effects, the value of $p$ is virtually unchanged, while the polarization angle
rotates a modest amount to $\theta = 83^{\circ}$.  The corrected value of
$\theta$ is nearly orthogonal to the arcsecond-scale radio and optical
emission-line structures observed along P.A.\ $\approx 0^{\circ}$ in
NGC 2110 (Gonz\'alez Delgado et al.\ 2002, and references therein).  This
is consistent with expectations based on the unified AGN model, whereby
scattering of the nuclear continuum occurs along a direction parallel to
the axis of the obscuring torus (Antonucci 1984).

A dramatic increase in polarization (peaking at $p \approx 5$\%) is observed
on either side of the H$\alpha$+[\ion{N}{2}] narrow-line blend in Figure~1.
As a result, the polarized flux spectrum of NGC 2110 ({\it bottom panel})
exhibits an extremely broad, double-peaked H$\alpha$ line.  The full-width
at zero-intensity of the broad H$\alpha$ line is roughly $2.7 \times 10^4$
km~s$^{-1}$.  A similar feature is not observed at H$\beta$, although the
H$\beta$ line is broader ($\sim 1200$ km~s$^{-1}$ FWHM) than the forbidden
lines in the polarized flux spectrum.  Previous studies have shown that the
polarized H$\alpha$ lines of hidden Seyfert~1s can be asymmetric and broad
(e.g., Schmidt et al.\ 2002), but because of the structure observed in its
H$\alpha$ profile, NGC~2110 is perhaps the most extreme example discovered
to date.

\section{NGC 2110 as a Double-Peaked Emission-Line AGN}

The broad H$\alpha$ line shown in Figure~1 closely
resembles those exhibited by double-peaked emission-line AGNs, a small
but important class of active nuclei (Eracleous \& Halpern 1994, 2003;
Strateva et al.\ 2003).  In fact, NGC 2110's broad H$\alpha$ profile is
nearly identical to that of the prototypical double-peaked AGN
Arp~102B (cf.\ Fig.~2 of Halpern et al.\ 1996).  An important secondary
characteristic of double-peaked emitters is that they often display strong
low-ionization forbidden lines, such as [\ion{N}{2}] $\lambda 6584$,
[\ion{O}{1}] $\lambda 6300$, and [\ion{S}{2}] $\lambda\lambda 6717,6731$.
Similar to many radio-selected double-peakers, the emission-line flux
ratios in NGC~2110 suggest a LINER classification (Heckman 1980; Veilleux
\& Osterbrock 1987): log ([\ion{N}{2}]/H$\alpha$) = 0.17,
log ([\ion{O}{1}]/H$\alpha$) = $-$0.23, and
log ([\ion{S}{2}]/H$\alpha$) = 0.00.
The Seyfert-like [\ion{O}{3}] $\lambda 5007$/H$\beta$ flux ratio of
$\sim 4$ in NGC~2110 is common among optically selected double-peaked
emitters (Strateva et al.\ 2003).  Thus, on spectroscopic grounds, NGC~2110
appears to be a genuine double-peaked emission-line AGN.  This marks the
first time that this type of object has been discovered via spectropolarimetry
in a Seyfert~2 galaxy.

Although at one time the origin of broad, double-peaked emission lines in AGNs
was a matter of considerable debate (see Gaskell et al.\ 1999), the prevailing
interpretation today is that such lines are produced in BLRs
that have a disk geometry (Eracleous \& Halpern 2003).  We have thus fitted
relativistic Keplerian disk models to the H$\alpha$ profile in the polarized
flux spectrum of NGC~2110.  A circular disk model (Chen \& Halpern 1989)
provides a rather poor fit, so we have instead fitted the line
profile with an elliptical disk model (following Eracleous et al.\ 1995)
under the assumption that the emissivity of the disk scales with radius
as $r^{-3}$ (Collin-Souffrin \& Dumont 1989; Eracleous \& Halpern 2003;
Strateva et al.\ 2006).  As illustrated in Figure~2, this
model provides a very good fit to the polarized H$\alpha$ line profile.
The best-fitting model is described by the following parameters: the inner
and outer pericenter distances of the line-emitting portion of the disk (in
units of the gravitational radius, $r_{\rm g}\equiv GM_{\rm BH}/c^2$, where
$M_{\rm BH}$ is the black-hole mass) are $\tilde\xi_1 = 200^{+110}_{-40}$
and $\tilde\xi_2 = 850^{+750}_{-250}$; the inclination of the disk axis
is $i=30^{\circ}\,^{+8^{\circ}}_{-4^{\circ}}$;
the eccentricity of the disk is $e=0.3\pm 0.1$; the orientation of the major
axis relative to the line of sight (see Fig.~2 of Eracleous et al.\ 1995) is
$\varphi_0=100^{\circ}\pm 30^{\circ}$; and the broadening parameter
(representing turbulent motions in the line-emitting ``skin'' of the disk)
is $\sigma=600^{+300}_{-200}~{\rm km~s}^{-1}$.  
These parameters are consistent with those obtained for Arp~102B by Chen \&
Halpern (1989), i.e., $\xi_1=350$, $\xi_2=1000$, $i=32^{\circ}$, and $\sigma=
850~{\rm km~s}^{-1}$.
(Note that because the broad H$\alpha$ line in NGC 2110 is reflected,
the angles $i$ and $\varphi_0$ describe the orientation of the disk to
the scattering medium, not the observer.)
The residuals of the fit, also shown in Figure~2, show a weak, broad,
bell-shaped pedestal underlying the H$\alpha$ + [\ion{N}{2}] complex.
Such a kinematic component is also observed in the H$\alpha$ and
Ly$\alpha$ profiles of Arp~102B (it is in fact the dominant component of the
Ly$\alpha$ line; Halpern et al.\ 1996). Based on a four-component Gaussian
fit to the H$\alpha$ + [\ion{N}{2}] lines in the residuals spectrum, we
estimate the width of the bell-shaped broad component to be
$\sim 3100~{\rm km~s}^{-1}$ (FWHM), similar to the widths of the
Ly$\alpha$ and H$\alpha$ lines of Arp 102B (Halpern et al.\ 1996).

For a final comparison to other double-peaked AGNs, we have estimated the
bolometric-to-Eddington luminosity ratio $L_{\rm bol}/L_{\rm Edd}$ of NGC~2110.
Using the Tremaine et al.\ (2002) $M_{\rm BH} - \sigma_*$ relation, the
observed stellar velocity dispersion of $\sigma_* = 220 \pm 25$ km~s$^{-1}$
(Nelson \& Whittle 1995) indicates a black-hole mass of $M_{\rm BH} = 2
\times 10^8$ $M_{\odot}$, which implies $L_{\rm Edd} \approx 3 \times
10^{46}$ ergs~s$^{-1}$.  The intrinsic 2--10 keV X-ray luminosity of $L_{\rm X}
\approx 4 \times 10^{42}$ ergs~s$^{-1}$ (Malaguti et al.\ 1999), combined
with a bolometric correction factor of $L_{\rm bol}/L_{\rm X} = 30$ (typical
of radio-quiet quasars; Elvis et al.\ 1994), yields $L_{\rm bol} \approx
1.2 \times 10^{44}$ ergs~s$^{-1}$.  We note that the
extinction-corrected [\ion{O}{3}] $\lambda 5007$ luminosity of $\sim 6
\times 10^{40}$ ergs~s$^{-1}$ derived from our Keck data, combined with a
bolometric correction of $L_{\rm bol}/L_{\rm 5007} = 3500$ (Heckman
et al.\ 2004), yields $L_{\rm bol} \approx 2.1 \times 10^{44}$ ergs~s$^{-1}$.
Thus, for NGC~2110 we estimate that $L_{\rm bol}/L_{\rm Edd} \approx 4
\times 10^{-3}$.  Despite the rough numbers used here, this is in close
agreement with the value of $L_{\rm bol}/L_{\rm Edd} =
(1-2) \times 10^{-3}$ obtained by Lewis \& Eracleous (2006) for
Arp~102B, which further strengthens the case that NGC~2110 is an obscured
twin of Arp~102B.

\section{Transient Nature of the Broad H$\alpha$ Line}

Close inspection of the top panel of Figure~1 reveals that the blue peak of
the broad H$\alpha$ line is faintly visible in the total-flux spectrum of
NGC~2110.  This is perhaps not too surprising, given the high polarization of
the feature and the unconfused nature of the spectrum in its vicinity.  To
investigate whether the double-peaked line was present in previous years, we
have combed the literature for published nuclear spectra of NGC~2110.  The
most recent spectrum we were able to find was one obtained on 30 December
2000 (UT) with the STIS instrument on board the {\it Hubble Space Telescope\/}
(Ferruit et al.\ 2004), which we have retrieved from the {\sl HST\/} archive.
A total exposure of 4445~s was obtained using a 0\farcs2 slit.
Our extraction of the spectrum covers a 5-pixel (0\farcs25) region centered
on the nucleus.  Before comparing the Keck and STIS data, we removed the
continuum from each spectrum.  To model the Keck continuum, we combined a
spiral galaxy bulge template (that of M31) with a power-law component.  The
continuum in the STIS spectrum, which contains little or no starlight because
of the small effective aperture, was removed by subtracting a constant.
We rescaled the continuum-subtracted STIS spectrum to match the strength of
the narrow H$\alpha$ line in the Keck spectrum.

The results of our comparison are displayed in Figure~3.  The narrow emission
lines exhibit somewhat different velocity widths and flux ratios in the Keck
and STIS data; this is a reflection of density stratification in the
narrow-line region of NGC 2110 (e.g., Filippenko \& Halpern 1984), which has
been sampled by different apertures.  The blue peak of the broad H$\alpha$
line, clearly visible in the Keck spectrum near 6425 \AA , is completely
absent in the high signal-to-noise ratio STIS spectrum.  Given the lower
amount of diluting starlight in the small 0\farcs2 $\times$ 0\farcs25 STIS
aperture, the blue peak would be at least as prominent in the STIS spectrum
if it were present.  We conclude, therefore, that the double-peaked H$\alpha$
line in the nucleus of NGC~2110 has appeared rather recently.  
Moreover, additional spectropolarimetric observations in 2007 February by
H.~Tran have shown that the line profile has varied significantly over an
interval of 14 months.
\footnote{see http://www.eso.org/sci/meetings/agnii2007/presentations/Tran.pdf}
The transient nature of this feature and its subsequent
variations are not without precendent; the emergence of double-peaked lines
has been noted in other AGNs (e.g., Pictor~A; Halpern \& Eracleous 1994;
Eracleous \& Halpern 1998), and numerous other objects exhibit variability
of the double-peaked line profile (e.g., Storchi-Bergmann et al.\ 1995;
Gezari et al.\ 2007).

\section{Conclusions}

Beginning with observations of NGC~1068 (Antonucci \& Miller 1985),
spectropolarimetry has played a crucial role in our understanding of the
physical nature of AGNs, and it continues to provide valuable insight.
Our results establish that NGC~2110 possesses a disk-like BLR, which expands
the range of properties exhibited by Seyfert~2s.  In addition, NGC~2110
presents some opportunities for interesting follow-up study.  As noted
above, the double-peaked feature is transient and variable, and it would be
worthwhile to monitor its behavior.  The observed variations may trace
changes in the structure of the BLR or in the properties of the scattering
medium.  Also, the similarity of NGC 2110's broad H$\alpha$ profile to that
of Arp 102B suggests that the line has not been distorted significantly in
the scattering process.  This has implications for the scattering geometry,
e.g., the scattering medium may only see the line-emitting disk over a narrow
range of inclination angles.  High-resolution imaging polarimetry and
detailed spectropolarimetric modeling (e.g., Smith et al.\ 2004, 2005)
would clarify this issue.  In the bigger picture, by
demonstrating that at least one double-peaked emitter contains the type
of nuclear obscuration believed to be present in most (if not all) classical
Seyfert galaxies, we can now extend the concept of unification to the
disk-emitting class of AGNs.  On parsec scales, double-peaked AGNs may
be structurally similar to other types of Seyfert nuclei.  If so, this
underscores the need to understand how physical differences might originate
in their broad-line regions.

\acknowledgments
The data presented herein were obtained at the W.~M.\ Keck Observatory, which
is operated as a scientific partnership among the California Institute of
Technology, the University of California, and NASA. The Observatory was made
possible by the generous financial support of the W.~M.\ Keck Foundation.
The work of A.~J.~B.\ was supported in part by the National Science Foundation
through grant AST-0548198.
M.~E.\ acknowledges the warm hospitality of the Department of Astrophysics at
the American Museum of Natural History.

\vskip 0.2 truein
\begin{figure}[h]
\begin{center}
\includegraphics[width=0.75\textwidth,angle=270.0]{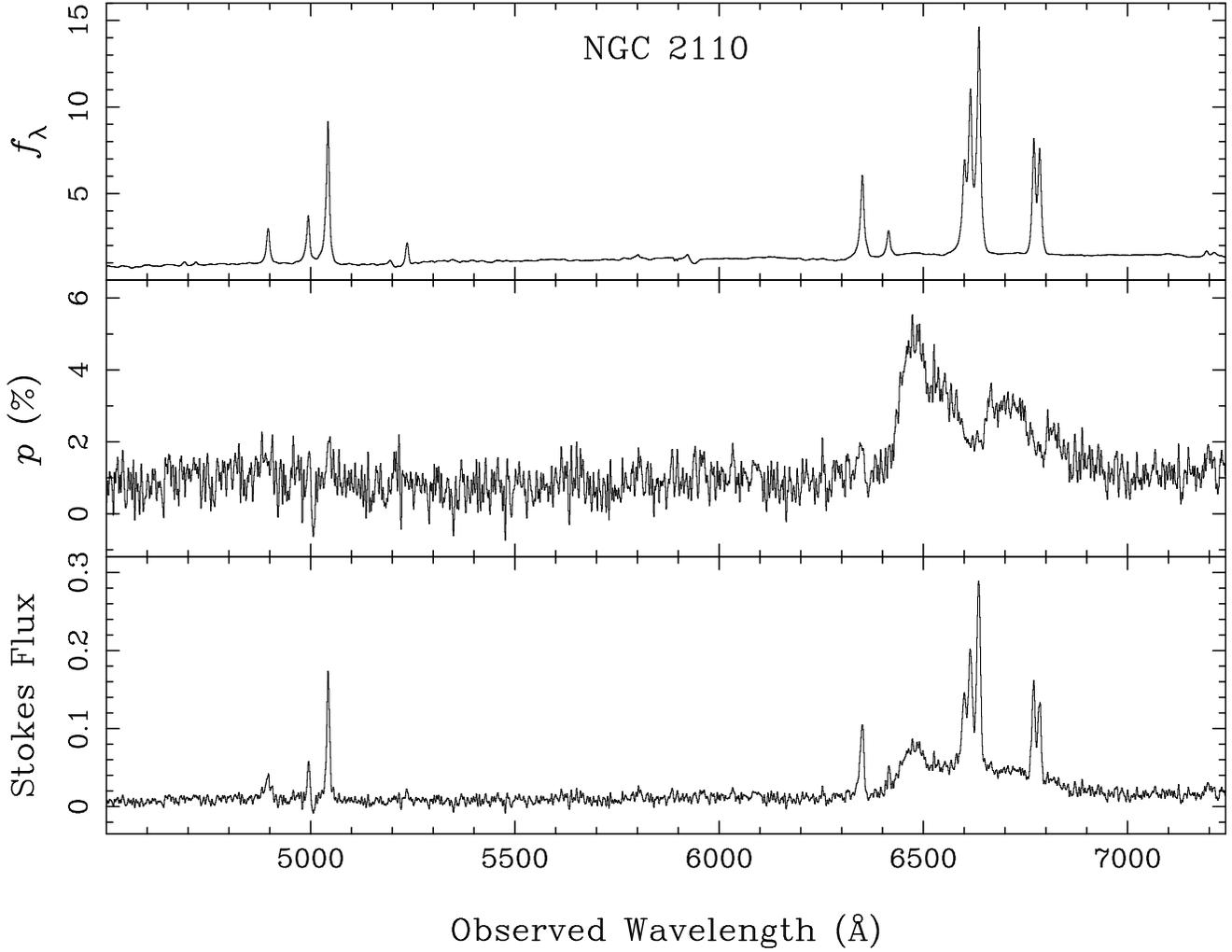}
\vskip 0.1truein
\caption{Spectropolarimetry of NGC~2110.  {\it Top}: Total flux, in
units of $10^{-15}$ erg cm$^{-2}$ s$^{-1}$ \AA$^{-1}$.  {\it Middle}:
Degree of linear polarization, given as the rotated Stokes parameter.
{\it Bottom}: Polarized flux, or ``Stokes flux,'' which is the product
of the total flux and rotated Stokes parameter.}
\end{center}
\end{figure}

\begin{figure}
\hbox to \hsize{\vbox{\hsize = 3.4truein
\includegraphics[width=0.37\textwidth,angle=270.0]{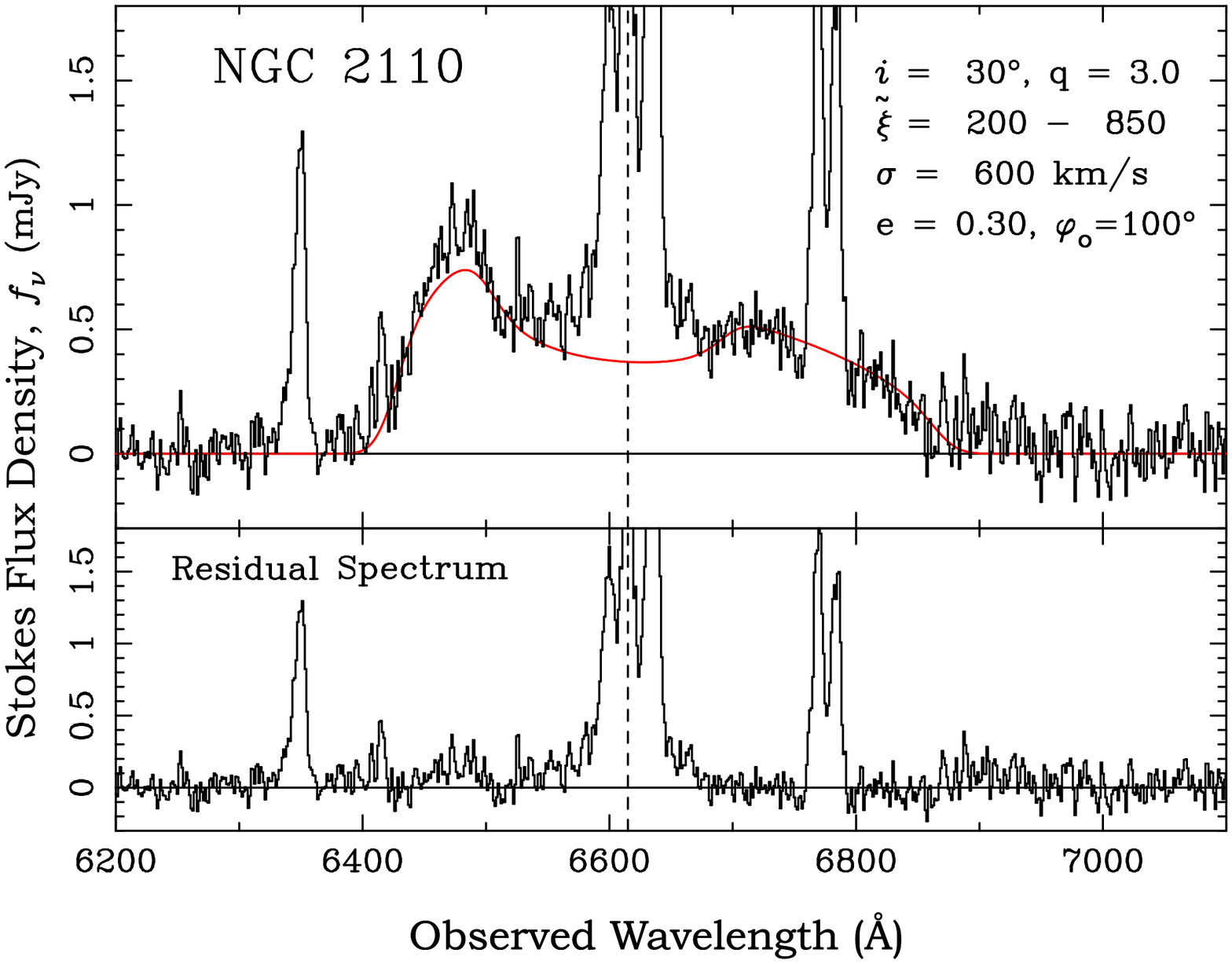}
\vskip 0.1truein
\caption{{\it Top panel}: The double-peaked H$\alpha$ line of NGC~2110, fitted
with an elliptical relativistic disk model.  See text for discussion of the
model parameters.  {\it Bottom panel}: Residuals from the fit.  Note the
additional weak, broad H$\alpha$ component that was not included in the
disk model.}
}}

\vskip 0.1truein
\hbox to \hsize{\vbox{\hsize = 3.4truein
\includegraphics[width=0.477\textwidth,angle=0.0]{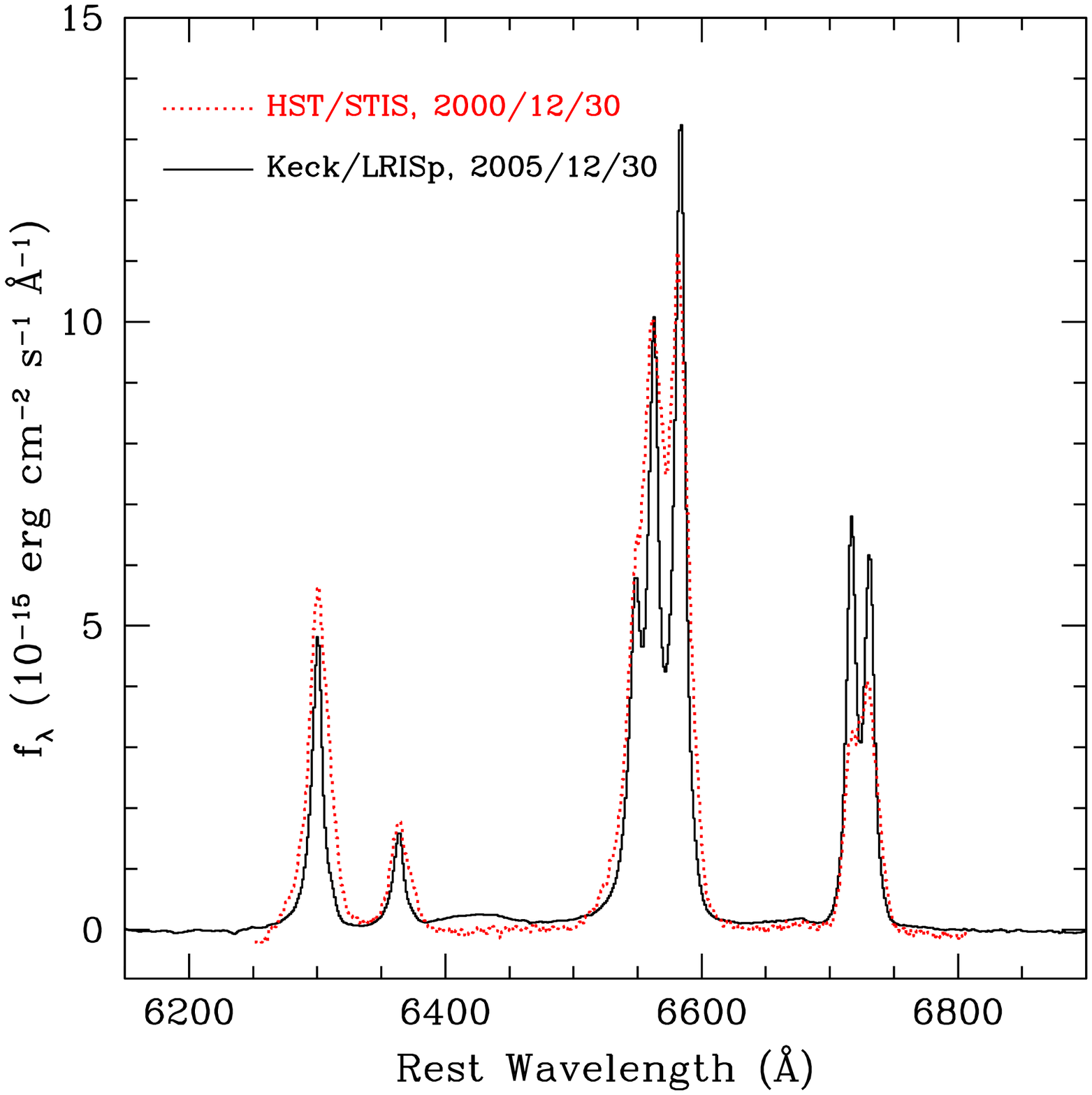}
\caption{Comparison of the total-flux spectrum of NGC~2110 from Keck with
an archival {\it HST}/STIS spectrum.  The absence in the STIS spectrum of
the $\sim$~6425 \AA\ bump present in the Keck data suggests that the
double-peaked H$\alpha$ line in NGC~2110 has emerged rather recently.}
}}
\end{figure}

\end{document}